\documentclass{article}
\usepackage{spconf,amsmath,graphicx}

\usepackage{amsthm,amsmath,amssymb}
\usepackage{mathrsfs}
\usepackage{subfig} 
\usepackage{diagbox}
\usepackage{booktabs}
 \usepackage{multirow}
\usepackage{amsmath}
\usepackage{float}

\title{AMC-Net: An Effective Network for Automatic Modulation Classification}
\name{Jiawei Zhang, Tiantian Wang, Zhixi Feng, Shuyuan Yang\thanks{This work was supported by the National Natural Science Foundation of China No.62276205.}}
\address{School of Artificial Intelligence, Xidian University, China}
\begin{document}

\maketitle

\begin{abstract}
Automatic modulation classification (AMC) is a crucial stage in the spectrum management, signal monitoring, and control of wireless communication systems. The accurate classification of the modulation format plays a vital role in the subsequent decoding of the transmitted data. End-to-end deep learning methods have been recently applied to AMC, outperforming traditional feature engineering techniques. However, AMC still has limitations in low signal-to-noise ratio (SNR) environments. To address the drawback, we propose a novel AMC-Net that improves recognition by denoising the input signal in the frequency domain while performing multi-scale and effective feature extraction. Experiments on two representative datasets demonstrate that our model performs better in efficiency and effectiveness than the most current methods.
\end{abstract}
\begin{keywords}
Automatic modulation classification, deep learning
\end{keywords}
\section{Introduction}
As wireless communication technologies advance rapidly, deep learning emerges as a potent technique for empowering wireless networks with complex topologies and radio conditions~\cite{mao2018deep}. Automatic modulation classification (AMC) as a crucial stage in communication systems, also has many combinations with deep learning~\cite{https://doi.org/10.48550/arxiv.2207.09647,https://doi.org/10.48550/arxiv.2205.01210}. Traditional AMC can be divided into two categories: $likelihood$-$based$ (LB) methods~\cite{panagiotou2000likelihood} and $feature$-$based$ (FB)
methods~\cite{moser2015automatic}. However, LB methods rely on prior knowledge about channel and signal. FB methods select hand-crafted features, then conduct the classification using the machine learning algorithm, such as support vector machines~\cite{li2017automatic} and random forests~\cite{triantafyllakis2017phasma}. FB methods highly depend on expert knowledge.
\\\indent O’Shea, Corgan, and Clanc~\cite{o2016convolutional} pioneer a CNN model for AMC, initiating the application of deep learning in AMC. It outperforms traditional methods that rely on manual features. A model based on LSTM is proposed by~\cite{rajendran2018deep}. Huang et al.~\cite{huang2020automatic} apply GRU to classify the signals. West and O’Shea~\cite{West2017DeepAF} apply a convolutional long short term deep neural network (CLDNN), which significantly improves classification accuracy. Moreover, the dual-stream structure of CNN-LSTM is proposed in~\cite{zhang2020automatic} to efficiently classify signals. It uses the information of I/Q channel and amplitude/phase to achieve better performance. I and Q channel are intergrated in~\cite{xu2020spatiotemporal} to learn the correlations of signals in parallel. Furthermore, Huynh-The et al.~\cite{huynh2020mcnet} implement different asymmetric convolution kernels and skip connections to learn spatial correlations. Liang et al.~\cite{liang2022radio} combine the attention mechanism and complex-valued neural network to better represent the signal. In addition to the focus on classification, AMC and deep learning have extended to other directions. Adversarial generative networks are introduced to AMC for data augment and better classification in~\cite{patel2020data}. Sahay, Brinton, and Love~\cite{sahay2021frequency} investigate the adversarial robustness in AMC. The above work extends the application of deep learning in AMC.
\\\indent In this paper, a novel modulation classification framework is proposed. This framework further considers the intrinsic properties of the modulated signals. It stands out from other frameworks in several special designs oriented to modulated signals. The contributions of this framework can be summarized as follow:
\\\indent $\bullet$ To mitigate the effects of noise and offset, we propose a novel \textit{Adaptive Correction Module} (ACM). By learning a set of weights in the frequency domain to correct the spectrum, it can eliminate noise in the original signal.
\\\indent $\bullet$ To learn multi-scale representations in modulated signals, we design a \textit{Multi-Scale Module} (MSM). It is composed of multiple convolutions with different receptive fields for capturing multi-scale features. It can effectively capture features of signal, such as amplitude, phase and frequency, at different scales.
\\\indent $\bullet$ For better learning of temporal correlation in signal sequences, we propose a \textit{Feature Fusion Module} (FFM) based on self-attention mechanism~\cite{vaswani2017attention}. It can handle long-distance dependence and support parallel computation.
\\\indent Experimental results show that the AMC-Net achieves state-of-the-art (SOTA) performance. In addition, our model is time-efficient compared with existing deep neural networks for AMC.
\section{methods}
\subsection{Problem Formulation}
Generally, the received signal $x_c(t)$ will be transformed into a discrete version $x_c[n]$ with a sampling rate $1 / T_{s}$. In practice, $x_c[n] \in \mathbb{C}$ is comprised of the in-phase component $I[n] \in \mathbb{R}$ and the quadrature component $Q[n] \in \mathbb{R}$. It can be described as $x[n]=I[n]+ j \cdot Q[n]$, where $j^2 = -1$. We denote $I[n]$ and $Q[n]$ as I and Q channel for short. By using I and Q channel, we express the $x_c \in \mathbb{C}^{1 \times L}$ as real-valued $x \in \mathbb{R}^{2 \times L}$. Overall, AMC aims to classify modulation format based on the $x \in \mathbb{R}^{2 \times L}$.

\subsection{Model Overview}
The overall architecture of AMC-Net is shown in Fig. 1. The input signal $x \in \mathbb{R}^{2 \times L}$ is passed through the ACM to obtain a more distinct representation. This is followed by the MSM, which integrates the information from the I and Q channel, and then three convolutional layers to extract deep spatial features. The extracted features are fed into the FFM for feature fusion in the temporal channel. Finally, results are fed into a classifier for classification after global average pooling. Next, we introduce the three modules in turn.

\begin{figure}[!b]
\centering

\includegraphics[width=0.5\textwidth]{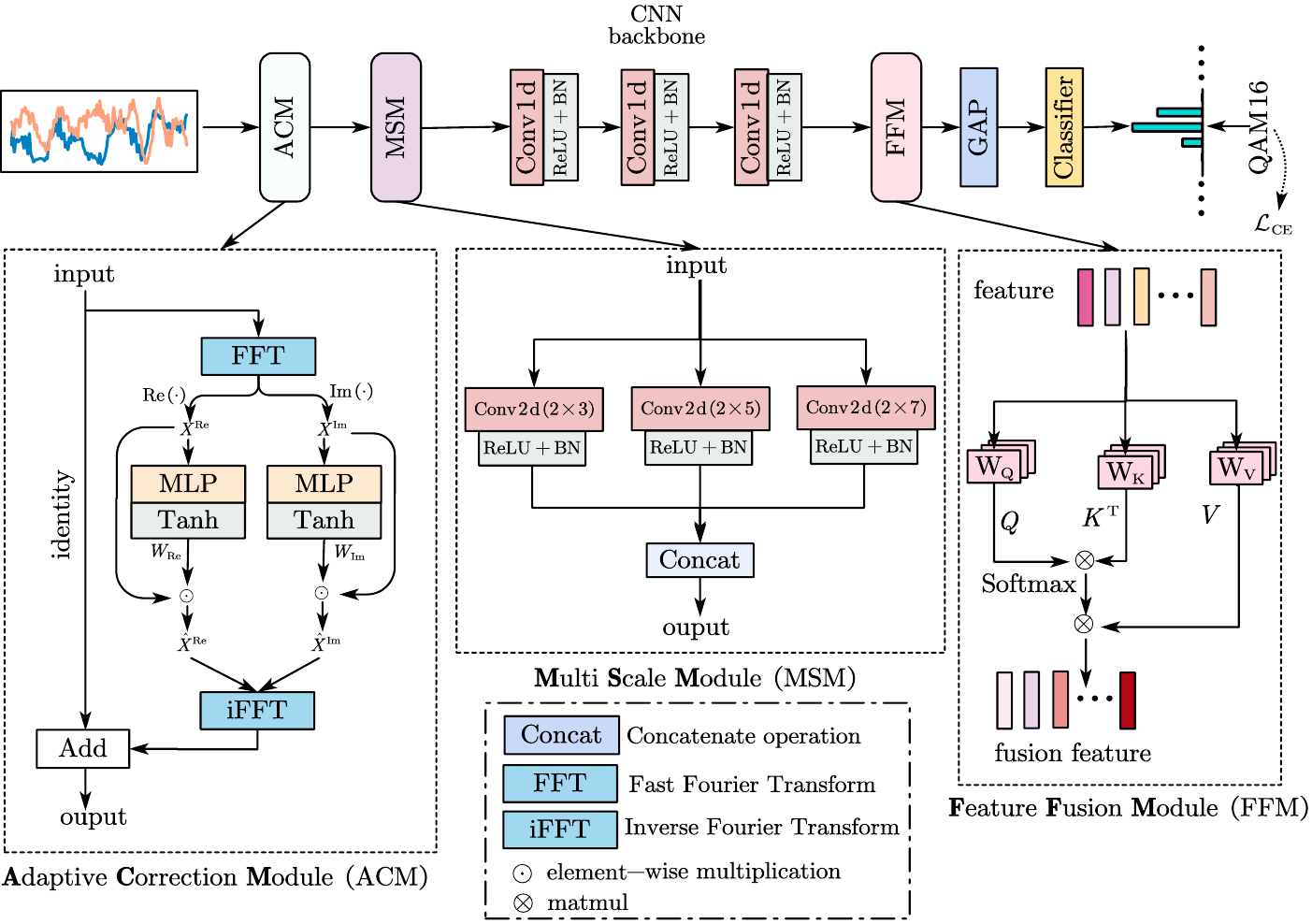}
\caption{The architecture of our AMC-Net. It consists of three parts, which are ACM, MSM, FFM.}
\end{figure}
\indent ACM essentially learns global filters from the frequency domain. Its intuition comes from the convolution theorem $\mathcal{F} \left\{ f\ast g \right\} =\mathcal{F} \left\{ f \right\} \cdot \mathcal{F} \left\{ g \right\}$. A complex-valued sequence $X_c \in \mathbb{C}^{1 \times L}$ is obtained by applying discrete Fourier transform to the input signal $x$. By using the real and imaginary parts, we express the $X_c \in \mathbb{C}^{1 \times L}$ as real-valued $X \in \mathbb{R}^{2 \times L}$. Namely:
\!
\begin{equation}
\begin{aligned}
\quad X_c[i] &=\sum_{k=0}^{L-1} (x[0,k] + j \cdot x[1,k]) e^{-j \frac{2 \pi}{L} i k} \\ 
\end{aligned}
\end{equation}

$X_c \in \mathbb{C}^{1 \times L} $ can be viewed as $X =\left[X^{\mathrm{Re}} ; X^{\operatorname{Im}}\right] \in \mathbb{R}^{2 \times L}$. The process of obtaining corrected spectrum $\hat{X}^{\mathrm{Re}}, \hat{X}^{\mathrm{Im}}$ can be expressed as:

\begin{center}
\begin{equation}
\begin{aligned}
&\hat{X}^{\mathrm{Re}}=\mathrm{Tanh} \left(\sigma_{\mathrm{Re}}\left(X^{\mathrm{Re}}\right)\right) \odot X^{\mathrm{Re}} \\
&\hat{X}^{\mathrm{Im}}=\mathrm{Tanh} \left(\sigma_{\mathrm{lm}}\left(X^{\mathrm{lm}}\right)\right) \odot X^{\mathrm{Im}}
\end{aligned}
\end{equation}
\end{center}
where $\sigma_{\mathrm{Re}}, \sigma_{\operatorname{Im}}$ are the mapping functions of two multilayer perceptron (MLP) respectively. We process the real and imaginary parts separately. Therefore, the changes in the real and imaginary parts are independent of each other, thereby increasing the adjustability of the spectrum in magnitude and phase. The tanh function restricts the output to the range $[-1,1]$, preventing the network from random projecting the values of the spectrum into an unconstrained range. This operation allows the processed signal to maintain the same order of magnitude as the original signal, making the network easier to train.

\indent After executing dot product in the spectrum, the inverse Fourier transform is applied to $\hat{X}$ to obtain the denoised signal $\hat{x}$. It can be expressed as:
\begin{equation}
\hat{x}[i]=\frac{1}{L} \sum_{k=0}^{L-1} \hat{X}[k] e^{j \frac{2 \pi}{L} i k}
\end{equation}
\\ \indent Skip-connection is deployed for associating $\hat{x}$ and $x$ to obtain the final output. The use of the skip-connection can help the network to retain signals and avoid severe nonlinear distortion from ACM. With the module we proposed, noise and offsets that interfere with classification can be attenuated to achieve better classification. 

The distinguishable differences for modulation formats usually appear on various scales than a single scale. It requires utilizing multi-scale features to capture those useful differences. Three convolutions are used in MSM, with convolution kernel sizes of 2 × 3, 2 × 5, and 2 × 7 respectively. The feature maps learned from different scales of the receptive field are concatenated to obtain a multi-scale feature map defined as
\begin{equation}
\begin{gathered}
X_{k}^{\prime}=\operatorname{ReLU}\left(\mathrm{BN}\left(\operatorname{Conv}_{k}(X)\right)\right), k \in\{1,2,3\} \\
X^{\prime \prime}=\operatorname{Concat}\left(X_{1}^{\prime}, X_{2}^{\prime}, X_{3}^{\prime}\right)
\end{gathered}
\end{equation}
where $X$ is the input vector, $\operatorname{Conv}_{\mathrm{k}}(\cdot)$ is the convolution with kernel size $(2,2 k+1)$, $X_{k}^{\prime}$ is the output of $\operatorname{Conv}_{\mathrm{k}}(\cdot),  X^{\prime \prime}$ is the multi-scale feature map.

The feature map is passed through three convolution layers to further extract high-level semantic features. The temporal features in high-level feature maps play a crucial role in predicting the modulation format. Therefore, we employ a encoder strucutre from multi-headed attention mechanism to fuse features. Multi-headed attention increases the representation subspace, enhancing the diversity and adaptability of feature representation. In contrast to previous work using LSTM, RNN, etc., self-attention enables the integration of information across the entire sequence, eliminating the problem of long-range dependence. Additionally, it can implement in parallel, thereby reducing inference times. The self-attention matrix is defined as:

\begin{equation}
\operatorname{Attention}(Q, K, V)=\operatorname{softmax}\left(\frac{Q K^{T}}{\sqrt{d_{k}}}\right) V
\end{equation}
where $Q \in \mathbb{R}^{d_{k} \times L}$ is queries, $K \in \mathbb{R}^{d_{k} \times L}$ is keys, $V \in \mathbb{R}^{d_{k} \times L}$ is values, $d_{k}$ is the queries and keys dimension, and $\sqrt{d_{k}}$ is the scaling factor for preventing excessive dot product values. And $Q, K, V$ are generated by linear projections. It can be expressed as:
\begin{equation}
Q=\mathbf{W}_{Q} X+\mathbf{b}_{Q}, K=\mathbf{W}_{K} X+\mathbf{b}_{K}, V=\mathbf{W}_{V} X+\mathbf{b}_{V}
\end{equation}
where $X\in\mathbb{R}^{C\times L}$ is the input feature map, $\mathbf{W}_{Q}\in\mathbb{R}^{d_{k} \times C},\\ \mathbf{W}_{K} \in \mathbb{R}^{d_{k} \times C}$, and $\mathbf{W}_{V} \in \mathbb{R}^{d_{k} \times C}$ are weight matrices for linear projections, $b_{Q}, b_{K}, b_{V} \in R^{d_{k}} $is the bias.

\indent The queries, keys, and values are independently projected $h$ times with different projection parameters, and the outputs of projections are concatenated to obtain the final output:
\begin{equation}
\text { MultiHead }=\text { Concat }\left(\text { head }_{1}, \ldots, \text { head }_{h}\right)
\end{equation}
where $\operatorname{head}_{i}=$ Attention $\left(Q^{i}, K^{i}, V^{i}\right)$, And MultiHead is the final output of FFM.

\section{Experimental Evaluation}
In this section, we evaluate the performance of the proposed AMC-Net on AMC task. And a thorough analysis of the experimental results is performed.
\subsection{Evaluation Datasets}
We evaluate on RML2016.10a and RML2016.10b~\cite{o2016convolutional} datasets. The two datasets are generated by GNU Radio. RML2016.10a comprises 11 modulation formats with SNR ranging from -20 to 18dB in 2 dB increments. This dataset has a total of 220000 examples, each modulation format has 1000 examples per SNR, and each example has 128 sampling points. Modulation formats in the dataset, including 8PSK, BPSK, QAM16, QAM64, QPSK, WBFM, CPFSK, GFSK, AM-DSB, AM-SSB, and PAM4, are widely used in modern wireless communication. RML2016.10b is an extended version of RML2016.10a. It contains 10 modulation formats except AM-SSB. And it has a total of 1.2 million examples. The two datasets contain actual channel defects, such as channel frequency offset, sampling rate offset, and noise to simulate imperfect transmission.

\subsection{Experimental Settings}

In our implement, the MLP in the ACM contains two fully connected layers of 128, 48, 128 nodes, activated by ReLU. The MSM has 12 filters for each different kernel size. The CNN backbone has three convolution layers with 64, 128, 256 filters and the size of filter is 1×3, followed by ReLU activation. The classifier contains two fully connected layers of 512, 256, 11/10 nodes. 2 headers are used in the FFM. Standard cross-entropy function serves as the loss function.
\\\indent Our model is trained by Adam optimizer with learning rate $\eta=10^{-3}$. For the two benchmark datasets, the training, validation, and test datasets (with a ratio of 6:2:2) are randomly selected from each modulation format in different SNRs. The default parameters are Xavier initialized for all layers. When the validation loss does not decrease in 10 epochs, the training process is stopped. All experiments are conducted using the PyTorch 1.8.1 library and a computer supported by NVIDIA CUDA with a GeForce GTX 2080Ti GPU.

\begin{table}[!p]
\centering
\caption{Comparisons of OA, macro-F1 and Kappa on RML2016.10a and RML2016.10b.}
\begin{tabular}{cccc} 
\toprule
\multirow{2}{*}{Model} & \multicolumn{3}{c}{RML2016.10a Dataset}                \\ 
\cline{2-4}
                       & OA                & macro-F1        & Kappa            \\ 
\hline
SVM                    & 18.55\%           & 0.1923          & 0.1472           \\
RF                     & 29.73\%           & 0.3023          & 0.2733           \\
MCLDNN                 & 61.23\%           & 0.6344          & 0.5776           \\
CCNN-Att               & 60.30\%           & 0.6243          & 0.5609           \\
Dual-Net               & 61.11\%           & 0.6328          & 0.5731           \\
MCNet                  & 57.38\%           & 0.5879          & 0.5360           \\ 
\hline
\textbf{AMC-Net}       & \textbf{62.51 \%} & \textbf{0.6483} & \textbf{0.5885}  \\ 
\hline
\multirow{2}{*}{Model} & \multicolumn{3}{c}{RML2016.10b Dataset}                \\ 
\cline{2-4}
                       & OA                & macro-F1        & Kappa            \\ 
\hline
SVM                    & 21.83\%           & 0.2214          & 0.1833           \\
RF                     & 32.46\%           & 0.3318          & 0.2841           \\
MCLDNN                 & 62.78\%           & 0.6338          & 0.5874           \\
CCNN-Att               & 63.03\%           & 0.6364          & 0.5901           \\
Dual-Net               & 64.05\%           & 0.6399          & 0.6017           \\
MCNet                  & 62.34\%           & 0.6279          & 0.5832           \\ 
\hline
\textbf{AMC-Net}       & \textbf{64.63 \%} & \textbf{0.6487} & \textbf{0.6081}  \\
\bottomrule 
\end{tabular}

\end{table}

\subsection{Baselines and Evaluation Metrics}
We compare with four SOTA models 
MCNet~\cite{huynh2020mcnet}, CCNN-Att~\cite{liang2022radio}, MCLDNN~\cite{xu2020spatiotemporal} and Dual-Net~\cite{zhang2020automatic}. And two traditional algorithms: support vector machine (SVM)~\cite{li2017automatic} and random forest (RF)~\cite{triantafyllakis2017phasma}.
\\\indent Following the common practices, we adopt overall accuracy, macro-averaged F1-score, and Kappa coefficient as the metrics in evaluating the overall classification performance.

\subsection{Main Results}

Table I displays the evaluation metrics for the two dataset experiments. Traditional machine learning algorithms can not adequately handle signals under complex communication situations, there is a significant difference between them and deep learning approaches. On both datasets, AMC-Net outperforms other SOTA models among deep learning techniques. This demonstrates the effectiveness of AMC-Net on AMC task.


\indent Table 2 shows the inference time and number of learned parameters for each model. The inference time counts the time each method takes to process a batch of signals. In general, AMC-Net achieves the best performance with faster inference at the expense of a small increase in the number of parameters, which is time-efficient.

\indent In Fig. 2, we plot the average accuracy at various SNRs for a more thorough analysis. AMC-Net not only outperforms all models at SNR less than 0 in RML2016.10a but also performs well at high SNR. The gap between the models begins to narrow in RML2016.10b due to an increase in the amount of data.

\begin{figure}[]
\centering 
    \subfloat[]{%
      \includegraphics[width=0.45\textwidth,height=0.27\textwidth]{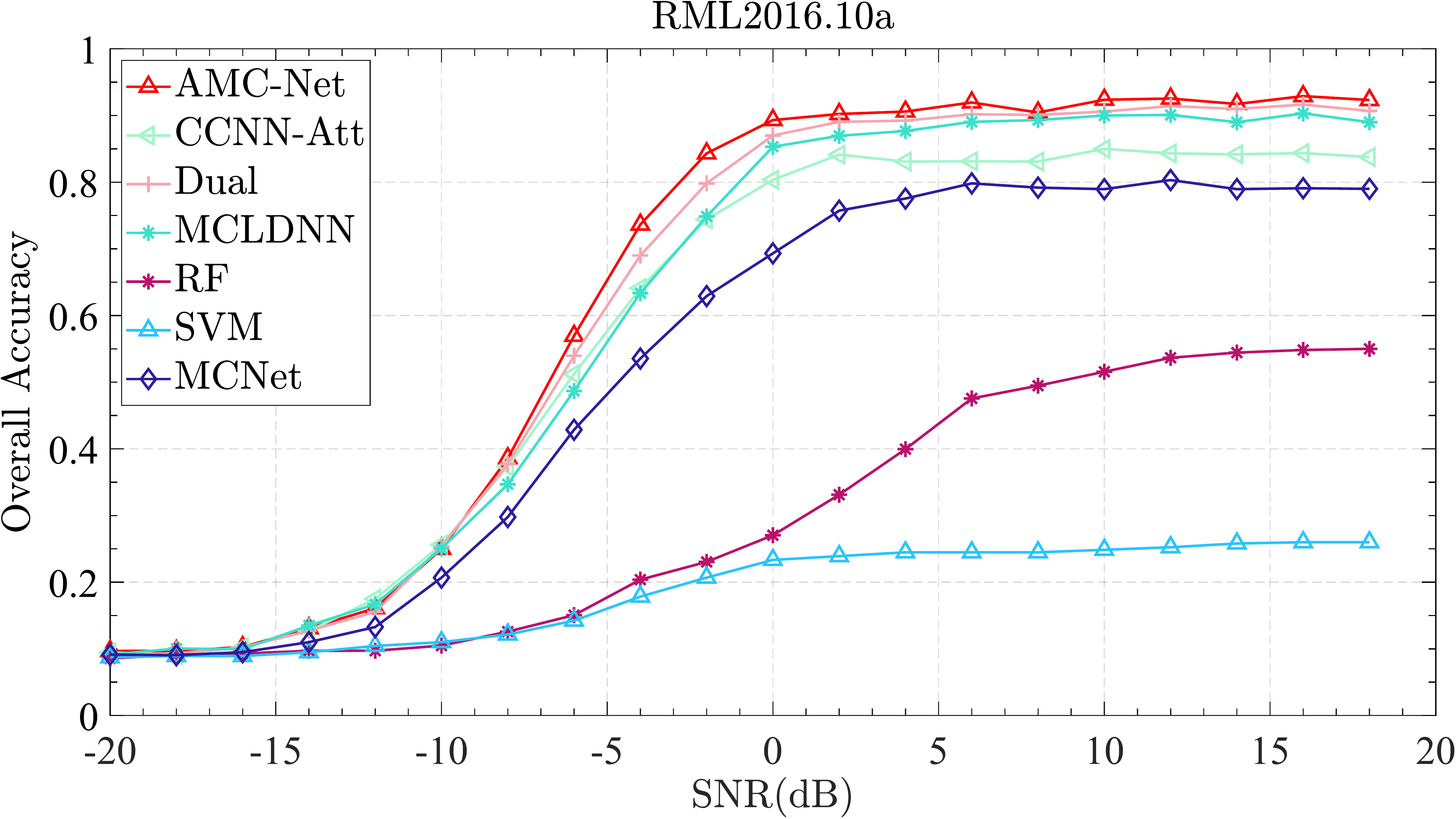}
    }
   \hspace{0.8in} 
    \subfloat[]{%
      \includegraphics[width=0.45\textwidth,height=0.27\textwidth]{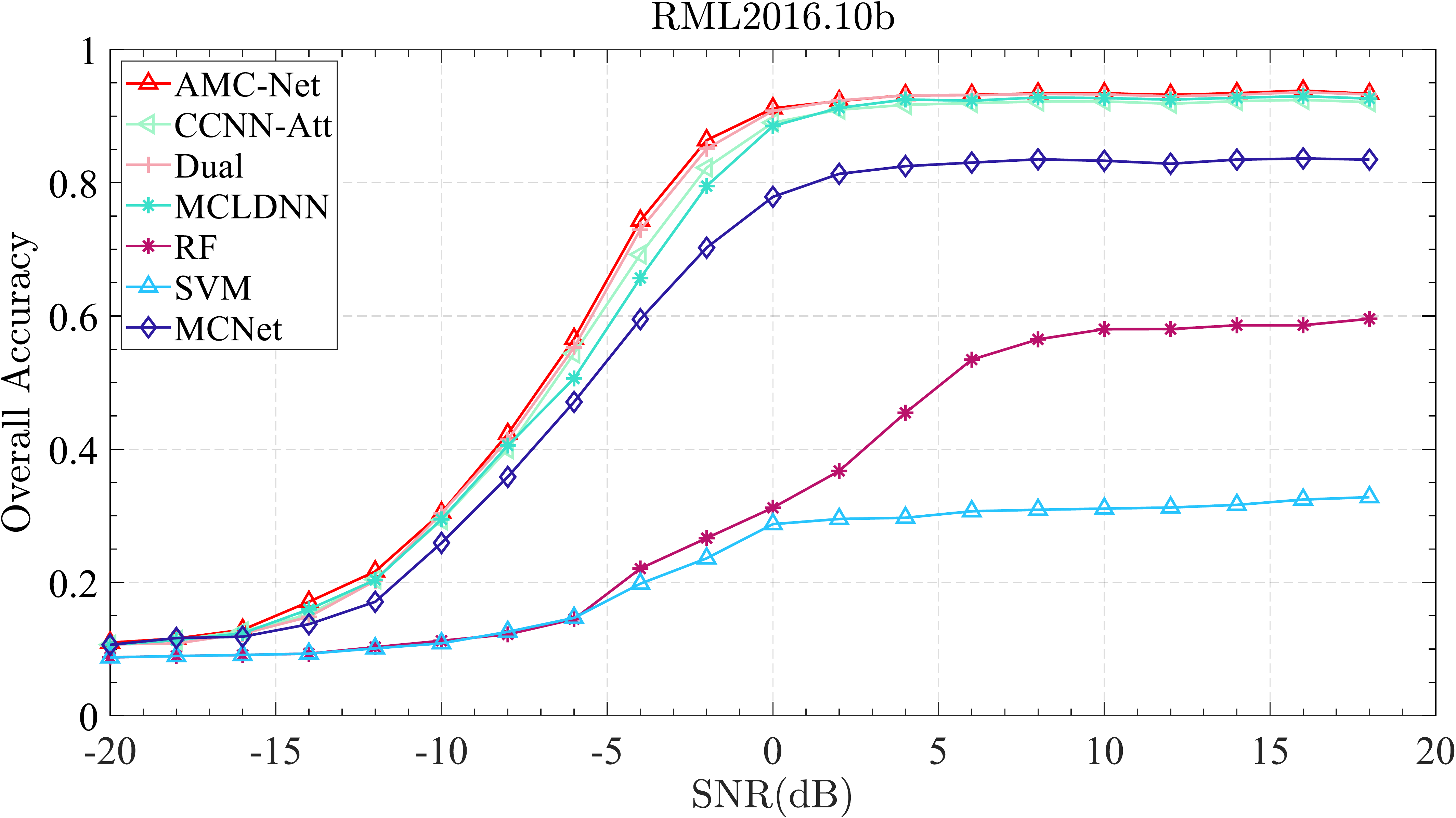}
    }
    \caption{Average accuracy comparisons of different SOTA methods on various SNR on the RML2016.10a (a), RML2016.10b (b).}
    \label{fig:dummy}
  \end{figure}

\begin{table}[!]
\centering
\caption{Learned parameters and inference time for the two datasets.}
\begin{tabular}{ccc} 
\toprule
Model            & \begin{tabular}[c]{@{}c@{}}Learned\\ Parameters\end{tabular}  & \begin{tabular}[c]{@{}c@{}}Inference\\ time(ms/batch)\end{tabular}  \\ 
\hline
SVM              & -                  & 0.018                                                                \\
RF               & -                  & 0.015                                                                \\
MCLDNN           & 0.41M              & 4.24                                                                 \\
CCNN-Att         & 0.38M              & 1.93                                                                 \\
Dual-Net         & 0.83M              & 9.12                                                                 \\
MCNet            & 0.13M              & 1.78                                                                 \\ 
\toprule
\textbf{AMC-Net} & 0.47M              & 1.52                                                                 \\
\bottomrule
\end{tabular}

\end{table}
\vspace{-0.3cm}
\subsection{Ablation Experiments}
We conduct ablation experiments to evaluate the contributions of the proposed modules.
\\\indent We remove MSM, FFM and ACM respectively, Table 4 shows the results of the ablation experiments. When the ACM is removed, we can see that the model accuracy drops the most, demonstrating its effectiveness. The effect of the FFM and the MSM slightly decreases in order. We claim it is because the ACM modifies the input signal distribution for subsequent networks to learn better. So when the ACM removed, the MSM and the FFM primarily concentrate on high-level feature extraction in a noisy signal distribution, which leads to the sub-optimal solution.
\begin{table}[]
\centering
\caption{OA, macro-F1, Kappa of the ablation experiments.}
\resizebox{0.4\textwidth}{!}{%
\begin{tabular}{cccc}
\toprule
model            & OA              & macro-F1        & Kappa           \\ \hline
AMC-Net w/o MSM  & 0.6135          & 0.6336          & 0.5749          \\
$\Delta$           & -1.16\%         & -1.47\%         & -1.36\%         \\ \hline
AMC-Net w/o ACM  & 0.6038          & 0.6277          & 0.5641          \\
$\Delta$           & -2.13\%         & -2.06\%         & -2.44\%         \\ \hline
AMC-Net w/o FFM  & 0.6093          & 0.6304          & 0.5702          \\
$\Delta$           & -1.58\%         & -1.79\%         & -1.83\%         \\ \hline
\textbf{AMC-Net} & \textbf{0.6251} & \textbf{0.6483} & \textbf{0.5885} \\ \bottomrule 
\end{tabular}%
}

\end{table}

\section{Conclusion}
In this paper, we propose a novel neural network for AMC task. To address the serious noise and offset interference during modulated signal transmission, our model employs an adaptive correction module in the frequency domain to reconstruct the signal. Meanwhile, a multi-scale module and a feature fusion module are proposed for efficient feature extraction and fusion. The combination of the three modules leads AMC-Net to a new SOTA level. Specifically, AMC-Net reveals the advantages of introducing frequency domain to time series, which may have the potential to be adopted for other time series classification tasks in the future.
\bibliographystyle{IEEEbib}
\bibliography{icassp.bib}
\newpage

\end{document}